\documentclass{article}
\usepackage{amsfonts}
\usepackage{amsmath}

\newtheorem{theorem}{Theorem}

\newtheorem{proposition}[theorem]{Proposition}

\newenvironment{proof}[1][Proof]{\noindent\textbf{#1.} }{\ \rule{0.5em}{0.5em}}
\input{tcilatex}
\begin{document}

\begin{center}
\bigskip

\bigskip

{\LARGE Bi-Hamiltonian Structure in Serret-Frenet Frame}

\bigskip

\bigskip

E Abado\u{g}lu and H. G\={u}mral

Department of Mathematics, Yeditepe University

Kay\i \c{s}da\u{g}\i\ 34750 \.{I}stanbul Turkey

\bigskip eabadoglu@yeditepe.edu.tr, \ \ hgumral@yeditepe.edu.tr

Nov 15, 2007
\end{center}

\bigskip

\textbf{Abstract }We reduced the problem of constructing bi-Hamiltonian
structure in three dimensions to the solution of a Riccati equation in
moving coordinates of Serret-Frenet frame. We then show that either the
linearly independent solutions of the corresponding second order equation or
the normal vectors of the moving frame imply two compatible Poisson
structures.

\bigskip

\section{Introduction}

The discovery of completely integrable nonlinear evolution equations as well
as the algebraic and geometric structures associated with them has triggered
an intensive search of finite dimensional dynamical systems resembling the
similar properties. The bi-Hamiltonian structure as an underlying
geometrical framework for complete integrability provoked the revival of
Poisson structures of finite dimensional dynamical systems (see \cite{olver}
and the references therein for details and comparison).

Several works \cite{nambu}-\cite{hojman} on construction of conserved
quantities, on Hamiltonian structures and on integrability of three
dimensional systems have led to a systematic investigation using Poisson
geometry, Frobenius integrability theorem and unfoldings of foliations \cite%
{phd}, \cite{hasan}. We presented correspondence between Poisson structures
and integrable one forms and utilized this to obtain criteria for local and
global existence of Poisson structures. We also obtained the local result
that any two Poisson structures can be made into a compatible pair to form a
bi-Hamiltonian structure.

The restrictions on Poisson matrices imposed by the Jacobi identity and the
compatibility condition for two of them to form a bi-Hamiltonian structure
are the most serious conditions requiring deliberate actions against their
simple presence as one scalar equation in three dimensions \cite%
{benitofairen}-\cite{ben}. In \cite{hasan}, using an invariance property, we
reduced the Jacobi identity to a nonlinear equation in ratios of components
of the Poisson matrix. This scalar equation in one unknown function was
shown to contain sufficient information for constructing the Poisson
structure completely and was recognized to be the Riccati equation in \cite%
{haas}.

The possibility to determine the Poisson matrix by a single function
straightens out the difficulties in general Hamiltonian systems which does
not fall into the classes of canonical Hamiltonian or Lie-Poisson (i.e. with
linear Poisson structure) equations, arising from the absence of coordinates
similar to the canonical Darboux coordinates of symplectic geometry. The
Darboux-Weinstein theorem \cite{weinstein} describes the local structure of
a Poisson manifold as a space foliated by symplectic submanifolds. The
foliation depends on the rank of the structure which is an invariant of the
Poisson matrix. As a result, the Poisson matrix consists of a constant
submatrix whose rank is the rank of the Poisson structure and some
additional nonlinear part. In three dimensions, one must solve for at least
one unknown function to determine the Poisson matrix completely.

The role of the constants of motion, in particular the Casimirs, in
linearization and integration of the equations associated with the Jacobi
identity were also discussed in \cite{benitofairen}-\cite{haas}. Yet, there
is no direct relations of the Jacobi identity and the compatibility
condition to the theory of linear differential equations which may be one of
the elegant ways to avoid some deceptive conclusions from these simple
looking differential equations ignoring their nonlinear character. The main
source of these confusions is to endeavor to exploit local criteria for
global results without questioning any suspicious obstructions such as the
one we have found for the Darboux-Halphen system \cite{hasan}.

In \cite{hasan} we also observed through several examples that \ some
coordinate transformations may cast the dynamical systems into a form where,
in spite of much higher degree of nonlinearity, the integration for the
conserved quantities and hence the manifestation of the bi-Hamiltonian
structures become more efficient. The non-covariance of the general
Hamiltonian formulation is a source of the common belief that the existence
of Hamiltonian structure and the integrability of dynamical systems rely
heavily upon the coordinates in which they are represented as well as the
prefered parametrization of the solution curves. Such coordinates were found
to be important in numerical integrations as well. It is shown in \cite%
{capel} that, a necessary condition for numerical solution algorithms to
preserve the conserved quantities of dynamical systems is their resemblance,
as products of a skew symmetric matrix and a gradient vector, to Hamiltonian
systems. See also \cite{bulent} for numerical schemes applied to some
concrete examples of systems under consideration. In three dimensions, the
Nambu mechanics \cite{nambu} is the only and generic (up to a conformal
factor) framework which enables us to identify such coordinates. Namely, the
Nambu structure is a manifestation of the bi-Hamiltonian structure, hence
integrability, in a frame with coordinate vectors consisting of the
dynamical vector field and gradients of two conserved Hamiltonians \cite%
{hasan}.

The Nambu representation obviously requires the integration of the system
for Hamiltonian functions. In our study of the Darboux-Halphen system, we
were able to obtain obstraction for the global integration of such
quantities \cite{hasan}. To our knowledge, this is the only example which
exhibits, along with a rich geometric structure, differentiation between
local criteria and global availability. On the other hand, the local version
of the Nambu mechanics has not yet been appeared in the literature.

In this work, we shall show that the coordinates associated with the
Serret-Frenet frame is the one we sought for three dimensional systems. In
these coordinates the Jacobi identity linearizes through the Riccati
equation and the compatibility follows from the Hamilton`s equations.
Obstructions to the global constructions of the bi-Hamiltonian structures
are encoded in the helicities and the cross-helicity of the unit vectors
spanning the normal plane to the vector field associated with the dynamical
system. We shall construct a bi-Hamiltonian moving frame for the local Nambu
representation.

In the next section we review the properties of bi-Hamiltonian systems in
three dimensions. In section three we introduce the Serret-Frenet frame
associated with a vector in three dimension. We express the Jacobi identity
in Serret-Frenet frame and show that in moving coordinates it reduces to a
Riccati equation. In section four, we shall show that Poisson structures
constructed from the solutions of the Riccati equations and/or the normal
vectors are all compatible via Hamilton`s equations of motion. We then
conclude the existence of bi-Hamiltonian structures.

\section{Hamiltonian Systems in Three Dimensions}

Following \cite{hasan}, we shall summarize the necessary ingradients of the
bi-Hamiltonian formalism in three dimensions. For $\mathbf{x=}\left\{
x^{i}\right\} =(x,y,z)\in 
%TCIMACRO{\U{211d} }%
%BeginExpansion
\mathbb{R}
%EndExpansion
^{3}$, $t\in 
%TCIMACRO{\U{211d} }%
%BeginExpansion
\mathbb{R}
%EndExpansion
$ and dot denoting the derivative with respect to $t$, we consider the
autonomous differential equations%
\begin{equation}
\overset{\cdot }{\mathbf{x}}=\mathbf{v}\left( \mathbf{x}\right)  \label{e1}
\end{equation}%
associated with a three-dimensional smooth vector field $\mathbf{v}.$This
equation is said to be Hamiltonian if the vector field can be written as 
\begin{equation}
\mathbf{v}\left( \mathbf{x}\right) =\Omega \left( \mathbf{x}\right) \left(
dH\left( \mathbf{x}\right) \right)  \label{e2}
\end{equation}%
where $H\left( \mathbf{x}\right) $ is the Hamiltonian function and $\Omega
\left( \mathbf{x}\right) $ is the Poisson bi-vector (skew-symmetric,
contravariant two-tensor) subjected to the Jacobi identity 
\begin{equation}
\left[ \Omega \left( \mathbf{x}\right) ,\Omega \left( \mathbf{x}\right) %
\right] =0  \label{e3}
\end{equation}%
defined by the Schouten bracket. In coordinates, if $\partial _{i}=\partial
/\partial x^{i}$ the Poisson bi-vector is $\Omega \left( \mathbf{x}\right)
=\Omega ^{jk}\left( \mathbf{x}\right) \partial _{j}\wedge \partial _{k}$
with summation over repeated indices. Then the Jacobi identity reads%
\begin{equation}
\Omega ^{i[j}\partial _{i}\Omega ^{kl]}=0  \label{e3c}
\end{equation}%
where $[jkl]$ denotes the antisymmetrization over three indices. It follows
that in three dimensions the Jacobi identity is a single scalar equation.
One can exploit the vector calculus and the differential forms in three
dimensions to have a more transparent understanding of Hamilton`s equations
as well as the Jacobi identity.\ Using the isomorphism 
\begin{equation}
J_{i}=\varepsilon _{ijk}\Omega ^{jk}\qquad i,j,k=1,2,3  \label{e4}
\end{equation}%
between skew-symmetric matrices and (pseudo)-vectors defined by the
completely antisymmetric Levi-Civita tensor $\varepsilon _{ijk}$ we can
write the Hamilton's equations $\left( \ref{e2}\right) $ in vector form 
\begin{equation}
\mathbf{v}=\mathbf{J}\times \nabla H  \label{e5}
\end{equation}%
and in this notation the Jacobi identity $\left( \ref{e4}\right) $ becomes 
\begin{equation}
\mathbf{J}\cdot \left( \nabla \times \mathbf{J}\right) =0  \label{e6}
\end{equation}%
In this form, the Jacobi identity is recognized to be equivalent to the
Frobenius integrability condition for the vector $\mathbf{J}$, or
equivalently, the condition for the one form $J=J_{i}dx^{i}$ to define a
foliation of codimension one in three dimensional space \cite{tondeur}, \cite%
{reinhart}, \cite{hasan}.

A distinguished property of Poisson structures in three dimensions is the
invariance of the Jacobi identity under the multiplication of the Poisson
vector $\mathbf{J}\left( \mathbf{x}\right) $ by an arbitrary but non-zero
factor. More precisely, one can easily show that under the transformation 
\begin{equation}
\mathbf{J}\left( \mathbf{x}\right) \rightarrow f(\mathbf{x})J\left( \mathbf{x%
}\right)  \label{e6a}
\end{equation}%
of Poisson vector the Jacobi identity transforms as%
\begin{equation}
\mathbf{J}\cdot \left( \nabla \times \mathbf{J}\right) \rightarrow \left( f(%
\mathbf{x})\right) ^{2}\mathbf{J}\cdot \left( \nabla \times \mathbf{J}\right)
\label{e6b}
\end{equation}%
which manifests the invariance property. The identities

\begin{equation}
\mathbf{J}\cdot \mathbf{v}=0,\text{ \ \ \ }\nabla H\cdot \mathbf{v}=0
\label{e11}
\end{equation}%
follows directly from the Hamilton's equations $\left( \ref{e5}\right) $,
the second of which is the expression for the conservation of the
Hamiltonian function.

A three dimensional vector $\mathbf{v}\left( \mathbf{x}\right) $ is said to
be bi-Hamiltonian if there exist two different compatible Hamiltonian
structures. In the notation of equation $\left( \ref{e5}\right) $, this
implies%
\begin{equation}
\mathbf{v}=\mathbf{J}_{1}\times \nabla H_{2}=\mathbf{J}_{2}\times \nabla
H_{1}  \label{e11a}
\end{equation}%
for the dynamical equations. The compatibility condition for $\mathbf{J}_{1}$
and $\mathbf{J}_{2}$ is defined by the Jacobi identity for the Poisson
vector $\mathbf{J}_{1}+c\mathbf{J}_{2}$ for arbitrary constant $c.$ Namely, $%
\mathbf{J}_{1}$ and $\mathbf{J}_{2}$ are compatible Poisson vectors provided
they satisfy%
\begin{equation}
\mathbf{J}_{1}\cdot \left( \nabla \times \mathbf{J}_{2}\right) +\mathbf{J}%
_{2}\cdot \left( \nabla \times \mathbf{J}_{1}\right) =0.  \label{e11b}
\end{equation}%
The invariance properties of the Jacobi identity and the Hamiltonian
functions enable one to extend the constant $c$ to be a function of the
conserved Hamiltonians. More precisely, the Jacobi identity for the
combination $\mathbf{J}_{1}+c\mathbf{J}_{2}$ of Poisson vectors gives

\begin{equation}
(\mathbf{J}_{1}\times \mathbf{J}_{2})\cdot \nabla c=(\mathbf{J}_{1}\cdot
\left( \nabla \times \mathbf{J}_{2}\right) +\mathbf{J}_{2}\cdot \left(
\nabla \times \mathbf{J}_{1}\right) )c  \label{e36}
\end{equation}%
which reduces to equation $\left( \ref{e11b}\right) $ whenever $c$ is a
constant. This linear equation can always be solvable for the function $c$
resulting in considerable relaxation in the compatibility condition. That
means, locally every pair of Poisson vectors can be made compatible.

It follows from the bi-Hamiltonian equations $\left( \ref{e11a}\right) $ and
the identities $\left( \ref{e11}\right) $ that $\mathbf{J}_{1}\times \nabla
H_{1}=\mathbf{J}_{2}\times \nabla H_{2}=0$. That is, the Hamiltonian of one
structure is the Casimir function of the other. Thus, a three dimensional
dynamical system can be defined to be integrable if it is a Hamiltonian
system with one Casimir. In this case, the flow can be represented by the
intersection of surfaces defined by constant values of the integrals of
motion (see \cite{hasan} and references below for further details and
examples).

\section{\protect\bigskip Jacobi Identity in Serret-Frenet Frame}

Let $\left( \mathbf{t},\mathbf{n},\mathbf{b}\right) $ denote the
Serret-Frenet frame associated with a differentiable curve $t\rightarrow 
\mathbf{x}(t)$ in some domain of the three dimensional space $%
%TCIMACRO{\U{211d} }%
%BeginExpansion
\mathbb{R}
%EndExpansion
^{3}$. Throughout, $\nabla =\left( \partial _{x},\partial _{y},\partial
_{z}\right) $ will denote the usual gradient operator in local Cartesian
coordinates. Given a vector field $\mathbf{v}$, the unit tangent vector $%
\mathbf{t},$ the unit normal $\mathbf{n},$ and the unit bi-normal $\mathbf{b}
$ can be constructed immediately as

\begin{equation}
\begin{array}{ccccc}
\mathbf{t}\left( \mathbf{x}\right) =\frac{\mathbf{v}\left( \mathbf{x}\right) 
}{\left\Vert \mathbf{v}\left( \mathbf{x}\right) \right\Vert } & \quad & 
\mathbf{n}\left( \mathbf{x}\right) =\frac{\mathbf{t}\times \left( \nabla
\times \mathbf{t}\right) }{\left\Vert \mathbf{t}\times \left( \nabla \times 
\mathbf{t}\right) \right\Vert } & \quad & \mathbf{b}\left( \mathbf{x}\right)
=\mathbf{t}\left( \mathbf{x}\right) \times \mathbf{n}\left( \mathbf{x}\right)%
\end{array}
\label{e15}
\end{equation}%
and they form a right-handed orthonormal frame except those vector fields $%
\mathbf{v}$ satisfying the condition imposed by 
\begin{equation}
\mathbf{t}\times \left( \nabla \times \mathbf{t}\right) =0.  \label{e15a}
\end{equation}%
It can be deduced from the vector identity $2\mathbf{t}\times (\nabla \times 
\mathbf{t)=\nabla (t\cdot t)+2t\cdot \nabla t=}$ $\mathbf{2t\cdot \nabla t}$
that this condition excludes essentially the flows with constant unit
tangent and the points $\mathbf{x}$ at which the unit normal $\mathbf{n}$
(hence the bi-normal $\mathbf{b}$) have zeros. That is, the cases one cannot
have a Serret-Frenet frame. To avoid this we may assume that 
\begin{equation}
\left( \nabla \times \mathbf{t}\right) \neq \lambda \left( \mathbf{x}\right) 
\mathbf{t}  \label{e15b}
\end{equation}%
\qquad \qquad \qquad for arbitrary nonzero function $\lambda \left( \mathbf{x%
}\right) $.\ That is, we exclude the dynamical systems whose unit tangent
vectors are the eigenvectors of the curl operator \cite{moses}, \cite{benn}.

We introduce the directional derivatives along the triad $\left( \mathbf{t},%
\mathbf{n},\mathbf{b}\right) $ as 
\begin{equation}
\begin{array}{ccccc}
\partial _{s}=\mathbf{t}\cdot \nabla & \quad & \partial _{n}=\mathbf{n}\cdot
\nabla & \quad & \partial _{b}=\mathbf{b}\cdot \nabla%
\end{array}
\label{e16}
\end{equation}%
so that the variables $(s,n,b)$ are the coordinates associated with the
Serret-Frenet frame. By inverting equations $\left( \ref{e16}\right) $ we
get the expression

\begin{equation}
\nabla =\mathbf{t}\partial _{s}+\mathbf{n}\partial _{n}+\mathbf{b}\partial
_{b}  \label{e17}
\end{equation}%
for the Cartesian gradient in Serret-Frenet frame. Since $\mathbf{t}\times
\nabla \times \mathbf{t=t\cdot \nabla t=\partial }_{s}\mathbf{t}$ the
definition of the normal vector reduces to one of the Serret-Frenet
equations justifying the name for the moving frame introduced \cite{andrade}.

It follows from the identity in equation $\left( \ref{e11}\right) $ that the
Poisson vector $\mathbf{J}$ has no component along the unit tangent vector $%
\mathbf{t}.$Hence, we set 
\begin{equation}
\mathbf{J}=\alpha \mathbf{n}+\beta \mathbf{b}  \label{e26}
\end{equation}%
for unknown functions $\alpha \left( \mathbf{x}\right) $ and $\beta \left( 
\mathbf{x}\right) $ satisfying $\alpha ^{2}+\beta ^{2}\neq 0$. Using
derivatives in Cartesian variables we find the expression%
\begin{equation}
\begin{array}{lll}
\mathbf{J}\cdot \left( \nabla \times \mathbf{J}\right) & = & \left( \beta
\nabla \alpha -\alpha \nabla \beta \right) \cdot \mathbf{t}+\alpha ^{2}%
\mathbf{n}\cdot \left( \nabla \times \mathbf{n}\right) +\beta ^{2}\mathbf{b}%
\cdot \left( \nabla \times \mathbf{b}\right) \\ 
&  & \text{ \ \ \ \ \ \ \ \ \ \ \ \ \ \ \ \ \ \ \ \ \ \ \ \ \ }+\alpha \beta
\left( \mathbf{n}\cdot \left( \nabla \times \mathbf{b}\right) +\mathbf{b}%
\cdot \left( \nabla \times \mathbf{n}\right) \right)%
\end{array}
\label{e28}
\end{equation}%
for the Jacobi identity. Assuming $\alpha \neq 0$ and defining the function $%
\mu =\beta /\alpha $ the Jacobi identity for $\alpha (\mathbf{n}+\mu \mathbf{%
b)}$ gives 
\begin{equation}
\mathbf{t\cdot }\nabla \mu =\mathbf{n\cdot }\nabla \times \mathbf{n}+\mu
\left( \mathbf{n\cdot }\nabla \times \mathbf{b}+\mathbf{b\cdot }\nabla
\times \mathbf{n}\right) +\mu ^{2}\mathbf{b\cdot }\nabla \times \mathbf{b}
\label{e30}
\end{equation}%
which is an equation involving only the unknown function $\mu $. Obviously,
this simplification is a manifestation of the invariance of the Jacobi
identity under the multiplication of $\ \mathbf{J}$ by an arbitrary but
non-zero function. Similarly, we may assume $\beta \neq 0$ and define $\eta
=-1/\mu =-\alpha /\beta $ for which the Jacobi identity for the combination $%
\beta (\mathbf{b}-\eta \mathbf{n)}$ becomes%
\begin{equation}
\mathbf{t\cdot }\nabla \eta =\mathbf{b\cdot }\nabla \times \mathbf{b}-\eta
\left( \mathbf{n\cdot }\nabla \times \mathbf{b}+\mathbf{b\cdot }\nabla
\times \mathbf{n}\right) +\eta ^{2}\mathbf{n\cdot }\nabla \times \mathbf{n.}
\label{e32}
\end{equation}

\bigskip\ To this end, we define the scalar quantities measuring the
non-integrability (in the sense of Frobenius) of each unit vector in the
Serret-Frenet triad%
\begin{equation}
\Omega _{\mathbf{t}}=\mathbf{t\cdot }\left( \nabla \times \mathbf{t}\right) ,%
\text{ \ }\ \Omega _{\mathbf{n}}=\mathbf{n\cdot }\left( \nabla \times 
\mathbf{n}\right) ,\text{ \ \ }\Omega _{\mathbf{b}}=\mathbf{b\cdot }\left(
\nabla \times \mathbf{b}\right)  \label{e33}
\end{equation}%
\ the first of which is necessarily not equal to any eigenvalue of the curl
operator by the assumption \ref{e15b}. The integration of these quantities
over the three space gives the so called helicities \cite{batch}, \cite%
{arnold} associated with the triad. We also introduce the sum

\begin{equation}
\Omega _{\mathbf{nb}}=\mathbf{n\cdot }\nabla \times \mathbf{b}+\mathbf{%
b\cdot }\nabla \times \mathbf{n}  \label{e33b}
\end{equation}%
of the cross-helicities for the normal and bi-normal vectors.

\begin{proposition}
Let $\mathbf{J}=\alpha \left( \mathbf{n}+\mu \mathbf{b}\right) $ (or $%
\mathbf{J}=\beta \left( \mathbf{b}-\eta \mathbf{n}\right) $). Then, the
Jacobi identity for $\mathbf{J}$ in moving coordinates is given by the
Riccati equation 
\begin{eqnarray}
\partial _{s}\mu &=&\Omega _{\mathbf{n}}+\mu \Omega _{\mathbf{nb}}+\mu
^{2}\Omega _{\mathbf{b}}  \label{e33ii} \\
\text{(or\ \ \ \ \ \ }\partial _{s}\eta &=&\Omega _{\mathbf{b}}-\eta \Omega
_{\mathbf{nb}}+\eta ^{2}\Omega _{\mathbf{n}}\text{ \ \ \ \ with \ \ }\mu
=-1/\eta \text{)}
\end{eqnarray}%
\qquad \qquad
\end{proposition}

Thus, in the moving coordinates, the Jacobi identity or, equivalently, the
existence of Poisson structure is expressible as a differential equation in
arclength coordinates only. It may be interesting to note that the equation
named after Jacopo Francesco Riccati originated from his investigations of
curves whose radii of curvature depend only on a single variable \cite{ince}%
. The disappearence of the moving coordinates $n$ and $b$ from the Jacobi
identity will become clear in the last section. In fact, they correspond to
local conserved quantities and, as discussed in \cite{hasan} for the
globally integrable cases, may appear arbitrarily in the Poisson vectors.

The Riccati equation $\left( \ref{e33ii}\right) $ is equivalent to the
linear second order equation 
\begin{eqnarray}
\partial _{ss}^{2}u-\left( \frac{\partial _{s}\Omega _{\mathbf{b}}}{\Omega _{%
\mathbf{b}}}+\Omega _{\mathbf{nb}}\right) \partial _{s}u+\Omega _{\mathbf{n}%
}\Omega _{\mathbf{b}}u &=&0\text{\ \ \ if }\ \Omega _{\mathbf{b}}\neq 0
\label{e34} \\
\text{(or \ }\partial _{ss}^{2}v-\left( \frac{\partial _{s}\Omega _{\mathbf{n%
}}}{\Omega _{\mathbf{n}}}-\Omega _{\mathbf{nb}}\right) \partial _{s}v+\Omega
_{\mathbf{n}}\Omega _{\mathbf{b}}v &=&0\text{ }\ \text{\ if }\Omega _{%
\mathbf{n}}\neq 0\text{\ )}
\end{eqnarray}%
with the solutions being related by%
\begin{equation}
\mu =-\frac{\partial _{s}\ln u}{\Omega _{\mathbf{b}}}\quad \text{\ if }\
\Omega _{\mathbf{b}}\neq 0\ \ \ \ \ \ \ \ \ \ \text{(or}\ \ \ \ \eta =-\frac{%
\partial _{s}\ln v}{\Omega _{\mathbf{n}}}\quad \text{\ if }\ \Omega _{%
\mathbf{n}}\neq 0\text{).}  \label{e35}
\end{equation}%
For the Poisson vectors of the above proposition at least one of the
equations in $\left( \ref{e34}\right) $ possesses two linearly independent
solutions.

The emergence of the Riccati equation as the Jacobi identity may be
interpreted as a relation between nonlinearity and superposition. The Jacobi
identity as well as the compatibility condition for Poisson vectors are
nonlinear restrictions on some linear combinations of basis vectors. Both
are scalar equations in three dimensions. The Riccati equation, on the other
hand, is known to be the only scalar equation admitting a nonlinear
superposition principle \cite{bountis}.

\section{\protect\bigskip Compatibility conditions}

To construct the bi-Hamiltonian structure, the Poisson vectors constructed
from the linearly independent solutions of the Riccati equation must be
compatible. That means, their linear combinations must also satisfy the
Jacobi identity. Although, the multiplicative factors are left arbitrary in
the construction of Poisson vectors they become important in the
compatibility condition. Apart from the general result discussed in section
two, we shall restrict ourselves to the case where $c$ is a constant in the
combination $\mathbf{J}_{1}+c\mathbf{J}_{2}$. We shall show that the
compatibility follows from the Hamilton`s equations.

First, we have the following result obtained by direct computation from the
compatibility condition

\begin{proposition}
Let $\alpha _{i}$ and $\mu _{i}$ be non zero and be different functions for $%
i=1,2$. For $\Omega _{\mathbf{b}}\neq 0$, the Poisson vectors $\mathbf{J}%
_{i}=\alpha _{i}(\mathbf{n+}\mu _{i}\mathbf{b)}$ are compatible if%
\begin{equation}
\partial _{s}\ln \frac{\alpha _{2}}{\alpha _{1}}=(\mu _{1}-\mu _{2})\Omega _{%
\mathbf{b}}.  \label{e37}
\end{equation}%
\ 
\end{proposition}

In obtaining Eq \ref{e37} we used the Riccati equations to eliminate
derivatives of functions $\mu _{1}$ and $\mu _{2}$. Next result shows that
above equation is always satisfied, via Hamilton`s equations, by Poisson
vectors constructed from the solutions of the Riccati equation.

\begin{proposition}
Let $\mathbf{J}=\alpha (\mathbf{n+\mu b)}$ and $H$ define a Poisson
structure for the dynamical system associated with $\mathbf{v}$. Then%
\begin{equation}
\partial _{s}\ln \frac{\parallel \mathbf{v\parallel }}{\alpha }-\mathbf{n}%
\cdot \nabla \times \mathbf{b}=\mu \Omega _{\mathbf{b}}  \label{e38}
\end{equation}

\begin{proof}
\ With the Poisson vector in the assumptions we write the dynamical system
as $\mathbf{t}=\parallel \mathbf{v\parallel }^{-1}\mathbf{J}\times \nabla
H=\parallel \mathbf{v\parallel }^{-1}\alpha (\mathbf{n+}\mu \mathbf{b}%
)\times \nabla H$. Taking cross products with $\mathbf{n}$ and $\mathbf{b}$
we get%
\begin{equation}
\mathbf{b=}\parallel \mathbf{v\parallel }^{-1}(\alpha \nabla H-\mathbf{%
J(n\cdot \nabla }H\mathbf{))}\text{ \ \ \ \ \ \ \ \ }\mathbf{n=-}\parallel 
\mathbf{v\parallel }^{-1}(\alpha \mu \nabla H-\mathbf{J(b\cdot \nabla }H%
\mathbf{))}  \label{e39}
\end{equation}%
from which we obtain $\nabla H=\alpha ^{=2}(1+\mu ^{2})^{=1}(\mathbf{%
J(J\cdot }\nabla H)+\mathbf{J}^{\perp }\parallel \mathbf{v\parallel }).$%
Here, we define $\mathbf{J}^{\perp }=\alpha (\mathbf{b-}\mu \mathbf{n)}$ for
convenience. The integrability condition $\nabla \times \nabla H=0$ for the
Hamiltonian function results, after taking dot product with $\mathbf{J,}$ in%
\begin{equation}
\partial _{s}\ln \frac{\parallel \mathbf{v\parallel }}{\alpha ^{2}(1+\mu
^{2})}=\frac{\mathbf{J}\cdot \nabla \times \mathbf{J}^{\perp }}{\alpha
^{2}(1+\mu ^{2})^{2}}  \label{e40}
\end{equation}%
where we used $\mathbf{J}\cdot \mathbf{J}^{\perp }=0$. The manipulations
leading to the result is now straightforward and requires only the use of
Jacobi identity for the derivative of $\mu $.
\end{proof}
\end{proposition}

The proof of compatibility becomes obvious once we write equation \ref{e38}
for each Poisson vector and subtract them. Similar results may be obtained
for the Poisson vectors in the form $\mathbf{J}=\beta \left( \mathbf{b}-\eta 
\mathbf{n}\right) $ by assuming $\Omega _{\mathbf{n}}\neq 0$ and using the
Riccati equation for the variable $\eta $.

We shall analyse the case $\Omega _{\mathbf{b}}=0$. This is the condition
for the unit vector $\mathbf{b}$ to satisfy the Jacobi identity. On the
other hand, $\Omega _{\mathbf{b}}=0$ reduces the Riccati equation to a
linear first order equation resulting in one linearly independent solution
for the construction of a Poisson vector of the form $\mathbf{J}=\alpha (%
\mathbf{n+}\mu \mathbf{b)}$.

\begin{proposition}
The compatibility condition for Poisson vectors $\mathbf{J}=\alpha (\mathbf{%
n+}\mu \mathbf{b)}$ and $\mathbf{b}$ is $\ \partial _{s}\alpha +\Omega _{%
\mathbf{nb}}=0$. This is satisfied by the Hamilton`s equations.

\begin{proof}
The equation can easily be obtained from the compatibility condition.
Equation \ref{e38} with $\Omega _{\mathbf{b}}=0$ holds for the Hamiltonian
structure for $\mathbf{J}$. For the Hamiltonian structure with the Poisson
vector $\mathbf{b}$, we have $\mathbf{t}=\parallel \mathbf{v\parallel }^{-1}%
\mathbf{b}\times \nabla H$ for some Hamiltonian function $H$. Taking
cross-product with $\mathbf{b}$, solving for $\nabla H$ and taking the dot
product of the equation resulting from the integrability condition $\nabla
\times \nabla H=0$, we find $\partial _{s}\mathbf{\ln }\parallel \mathbf{%
v\parallel =-b}\cdot \nabla \times \mathbf{n}$ which yields the result.
\end{proof}
\end{proposition}

Thus, the Poisson vectors obtained from solutions of Riccati equations are
always compatible.

\section{\protect\bigskip Bi-Hamiltonian structure}

We shall combine the results of the previous two sections on construction of
Poisson vectors and their compatibility to present the main result. This
will include the remaining case where the Poisson vectors are defined by the
normal and bi-normal unit vectors of the Serret-Frenet triad. In connection
with this particular structure we shall first relate the present work to the
existing examples of bi-Hamiltonian dynamical systems in the literature and
then present the local form of the Nambu mechanics.

\begin{proposition}
Every three dimensional dynamical system possesses two compatible Poisson
vectors.

\begin{proof}
If both $\Omega _{\mathbf{n}}$ and $\Omega _{\mathbf{b}}$ are non-zero, then
any of the two Riccati equations which produce the same result by
construction, give two Poisson structures coming from the linearly
independent solutions of the corresponding second order equation. If $\Omega
_{\mathbf{b}}=0$ (or $\Omega _{\mathbf{n}}=0$) then the first (the second)
Riccati equation becomes linear with one linearly independent solution. Note
that the sum $\Omega _{\mathbf{nb}}$ of the cross-helicities of the normal
vectors involves as an integrating factor in the integration for this
Poisson structure. The other Poisson structure is defined by the bi-normal
vector $\mathbf{b}$ (or the normal vector $\mathbf{n}$) since $\Omega _{%
\mathbf{b}}=0$ (or $\Omega _{\mathbf{n}}=0$) is the Jacobi identity for
this. The compatibility conditions for these cases are shown to be satisfied
via Hamilton`s equations in the previous section. If, we have both $\Omega _{%
\mathbf{n}}\mathbf{=}0$ and $\Omega _{\mathbf{b}}=0$ then $\mathbf{n}$ and $%
\mathbf{b}$ satisfies Jacobi identity and they become Poisson vectors we
sought. The compatibility condition is $\Omega _{\mathbf{nb}}=0$. This
implies that the function $\mu $ (or $\eta $) must be a non-zero constant in
the vector $\mathbf{J}$ for which the Riccati equation is now trivial. The
Hamilton`s equations with the Poisson vectors $\mathbf{n}$ and $\mathbf{b}$
implies $\partial _{s}\mathbf{\ln }\parallel \mathbf{v\parallel =n}\cdot
\nabla \times \mathbf{b}$ and $\partial _{s}\mathbf{\ln }\parallel \mathbf{%
v\parallel =-b}\cdot \nabla \times \mathbf{n}$, respectively. Eliminating
the vector $\mathbf{v}$, we obtain the compatibility condition $\Omega _{%
\mathbf{nb}}=0$.
\end{proof}
\end{proposition}

The last case of the theorem gives clues to relate the present work to the
existing bi-Hamiltonian dynamical systems in the literature, in particular,
to the Nambu mechanics. Recall that the three dimensional dynamical systems
admitting bi-Hamiltonian structure with $(\mathbf{J}_{1},H_{2})$ and $(%
\mathbf{J}_{2},H_{1})$ are of the form of equation \ref{e11a} with $\mathbf{J%
}_{1}$ and $\mathbf{J}_{2}$ satisfying Frobenius integrability conditions as
Jacobi identities. In case that these vectors are globally integrable they
are related to the conserved Hamiltonians by $\mathbf{J}_{i}=\varphi
_{i}\nabla H_{i}$ $i=1,2$ for some arbitrary non-zero functions $\varphi
_{i} $. In this case, the dynamical system has the form $\mathbf{v=\psi }%
\nabla H_{1}\mathbf{\times }\nabla H_{2}$ first studied by Nambu in \cite%
{nambu}. All explicitly constructed bi-Hamiltonian systems in three
dimension has this form \cite{nutku}, \cite{nutk}, \cite{hasan}, \cite{aygur}%
. The flow lines coincide with the intersections of the surfaces defined by
contant values of the Hamiltonians. Any constant appearing in this
bi-Hamiltonian picture can be taken as arbitrary functions of $H_{1}$ and $%
H_{2}$.

Our aim is to find the local version of this bi-Hamiltonian representation
in three dimensions. This we shall do by first proving the more general
result that the local existence of a Hamiltonian function of the proposed
form is equivalent to the existence of a Poisson vector (see also \cite{bben}%
).

\begin{proposition}
\bigskip There exists non-zero functions $H$ and $\varphi $ with $\mathbf{J=}%
\varphi \nabla H$ whenever $\mathbf{J=n+\mu b}$ satisfies the Jacobi
identity.

\begin{proof}
The condition $\nabla \times \nabla H\equiv 0$ together with the form of $%
\mathbf{J}$ gives $\nabla \times (\mathbf{n+\mu b)=a}\times (\mathbf{n+\mu b)%
}$ where $\mathbf{a}=\nabla \ln \sqrt{1+\mu ^{2}}/\varphi $. Taking dot
products with the unit vectors of the Serret-Frenet triad we obtain three
equations one of which is algebraic and the other two expressing the $s$ and 
$n$ derivatives of the function $\mu $. All three contains terms involving
the vector $\mathbf{a}$. Eliminating this term for the $s$ derivative
results in the Riccati equation \ref{e33ii} for the Poisson vector $\mathbf{J%
}$. The remaining two equations determine the function $\varphi $ and the
dependence of $\mu $ on the variable $n$.
\end{proof}
\end{proposition}

Thus, finding two linearly independent solutions of the Ricatti equation
completely determines the local bi-Hamiltonian structure.

In the local picture of the present work the normal coordinates $n$ and $b$
represents the local conserved quantities and they appear arbitrarily in the
Poisson structures. The normal vectors $\mathbf{n}$ and $\mathbf{b}$
defining the bi-Hamiltonian structure in the last case thus corresponds, in
the globally integrable case, to the gradients of Hamiltonian functions
defining Poisson vectors. Therefore, the manifestly bi-Hamiltonian equation $%
\mathbf{t=n\times b}$ corresponding to the last case of the above
proposition is the local version of the Nambu structure.

\bigskip

\bigskip

\bigskip

\end{document}